\documentclass[aip,10pt]{revtex4}
\usepackage{amsmath}
\usepackage{graphicx}

\begin{document}

\title{Jacobi $\vartheta $-functions and Discrete Fourier Transforms}
\author{M. Ruzzi}
\affiliation{Instituto de F\'{\i}sica Te{\'o}rica - S\~ao Paulo State University \\
01405 - 900 Rua Pamplona, 145 \\
S\~ao Paulo, SP, Brazil}

\begin{abstract}
Properties of the Jacobi $\vartheta _{3}$-function and its derivatives under
discrete Fourier transforms are investigated, and several interesting
results are obtained. The role of modulo $N$ equivalence classes in the theory
of $\vartheta $-functions is stressed. An important conjecture is studied.
\end{abstract}

\maketitle

\section{Introduction}

Methods in mathematical physics usually provide an interface between quite
different areas of physics, and it is not unusual that such areas advance in
parallel, mostly ignoring the each other's steps. This is the case with
finite dimensional inner product spaces (hereafter mentioned as the
``discrete''), with its leading role in Quantum Mechanics (hence Quantum
Information Theory) and in finite signal analysis. References \cite%
{vourdas,natig,vourdas2} provide some links between those theories).

Both quantum mechanics of finite dimensional Hilbert spaces and finite
signal analysis rely heavily on the Discrete Fourier Transform (DFT,
sometimes mentioned Finite or Fractional Fourier Transform), and, regarding
Quantum Mechanics, after the seminal work of Weyl on finite dimensional
systems \cite{weyl}, it was Schwinger who observed and explored the fact
that two physical observables whose families of eigenstates are connected
\textit{via }DFT share a maximum degree of incompatibility \cite{schw}.

Although, at first glance, a finite system might look much simpler than
anything defined on a non enumerable infinite dimensional Hilbert space
(hereafter referred to as the\ ``continuum''), there is much more knowledge
about the later than the former. In one phrase, in the continuum we have
one, and only one, harmonic oscillator, while in the discrete there are a
lot of candidates for that role, each one surely with its virtues, but
surely no undisputed champion.

The eigenstates associated to the harmonic oscillator, the Gaussian function
and the Hermite polynomials, have a very distinguishable behavior under the
action of the (usual) Fourier transform, so widely known that any comment on
this regard is completely superfluous. Over such properties rests a huge
amount of physical knowledge. On the other hand, however, although the
Discrete Fourier Transform (DFT) is a well known tool, there is nothing on
this context which could claim for itself a role analogous to that of the
Gaussian function/Fourier transform ``duo''.

A decisive step in an attempt to \ \textquotedblleft
regain\textquotedblright, in the discrete, all interpretative power derived
from the qualitative behavior of the harmonic oscillator eigenfunctions,
lost when one leaves the continuum realm, was given in \cite{mehta}, where
the eigenstates of the DFT are obtained. The purpose of this paper is to
further explore this path, showing new results which closely parallel those
of the continuum. Those results are obtained in a strikingly simple fashion,
exploring the technique of breaking infinite sums in modulo $N$ equivalence
classes. Pertinent research on the eigentates of the DFT can also be found in \cite{DFT}.

A remark must be made about the orthogonality of the DFT's eigenstates.
Mehta has conjectured that those states are indeed orthogonal, what seems to
be most reasonable. One may be led to believe that, just as in the
continuum, the eigenstates of the DFT may be also (non-degenerate)
eigenstates of some other (unitary or self-adjoint) operator, and thus
orthogonal. Further evidence supporting such conjecture is that the
continuous limit of the DFT eigenstates recovers, as expected, the Gaussian
times the Hermite polynomials. However, as it will be shown, quite
surprisingly, the conjecture \emph{does not} hold, giving another fine
example of the peculiarities of the finite dimensional context.

The eigenstates of the DFT are seen to be the Jacobi $\vartheta _{3}$%
-function and its derivatives \cite{gama}. Interest in Jacobi $\vartheta $%
-functions, by their own turn, may come from a variety of directions. First,
its mathematical interest goes without saying (see, for example, \cite{bell}
and references therein). To cite relatively recent examples in physics, in
quantum physics it is deeply related to coherent states associated to both
circle \cite{circle} and finite lattice topology \cite{nois}. Its modular
properties have proven to be of fundamental importance in superstring
theory, as it is shown by standard literature in this field \cite{cordas}.

The basic notation adopted in this article and some preliminary results are
presented in the next section. Following, orthogonality of the DFT's
eigenstates is discussed. Section IV contains the main results, for which a
two variable generalization is verified in the subsequent section. Further
relations among $\vartheta _{3}$ functions are obtained in section VI, which
precedes the concluding section.

\section{Preliminary results}

In reference \cite{mehta} it is shown that there is a set of functions with
the following remarkable property
\begin{equation}
f_{n}(j)=\frac{i^{n}}{\sqrt{N}}\sum_{k=0}^{N-1}f_{n}(k)\exp \left[ \frac{%
2\pi i}{N}kj\right] ,  \label{mehta}
\end{equation}%
where $N$ is a natural number. The functions
\begin{equation}
f_{n}(j)=\sum_{\alpha =-\infty }^{\infty }\exp \left[ -\frac{\pi }{N}(\alpha
N+j)^{2}\right] H_{n}\left( \epsilon (\alpha N+j)\right) ,\qquad \epsilon =%
\sqrt{\frac{2\pi }{N}}  \label{def1}
\end{equation}%
are defined making use of the Hermite polynomial $H_{n}$. Writing $H_{n}(x)$
in terms of its generating function, $H_{n}\left( x\right) =\left. \frac{%
\partial ^{n}}{\partial t^{n}}\exp \left[ 2xt-t^{2}\right] \right\vert
_{t=0},$ it is possible to write this state (to use a quantum mechanical
terminology) as \cite{gama}
\begin{equation}
f_{n}(j)=\frac{1}{\sqrt{N}}\left. \frac{\partial ^{n}}{\partial t^{n}}%
\vartheta _{3}\left( \frac{j}{N}-\frac{\epsilon }{\pi }t,\frac{i}{N}\right)
\exp \left[ t^{2}\right] \right\vert _{t=0},  \label{def2}
\end{equation}%
where
\begin{equation}
\vartheta _{3}\left( z,\tau \right) =\sum_{\alpha =-\infty }^{\infty }\exp
\left[ i\pi \tau \alpha ^{2}\right] \exp \left[ 2\pi i\alpha z\right]
,\qquad \mbox{Im}(\tau )>0,  \label{theta}
\end{equation}%
is the Jacobi $\vartheta _{3}$-function, following Vilenkin's notation \cite%
{vil}. In this notation the basic properties of this even function read as
\begin{eqnarray}
\vartheta _{3}\left( z+m+n\tau ,\tau \right)  &=&\exp \left[ -i\pi \tau n^{2}%
\right] \exp \left[ -2\pi inz\right] \vartheta _{3}\left( z,\tau \right)
\label{prop1} \\
\vartheta _{3}\left( z,i\tau \right)  &=&\tau ^{-1/2}\exp \left[ -\frac{\pi
z^{2}}{\tau }\right] \vartheta _{3}\left( \frac{z}{i\tau },\frac{i}{\tau }%
\right) ,  \label{prop2}
\end{eqnarray}%
emphasizing its period $1$ and quasi-period $\tau .$ A beautiful consequence
of (\ref{prop2}) is that this function can be written as a sum of Gaussians
\begin{equation}
\vartheta _{3}\left( \frac{z}{L},\frac{i}{\sigma ^{2}}\right) =\sigma
\sum_{\alpha =-\infty }^{\infty }\exp \left[ -\pi \left( \frac{\sigma }{L}%
\right) ^{2}(\alpha L+z)^{2}\right] ,  \label{gaussum}
\end{equation}%
a form in which the width $\frac{L}{\sigma }$ becomes apparent. Property (%
\ref{prop2}) also provides an easy way to obtain the additional identity
(also given by \cite{mehta})
\begin{equation}
f_{n}(j)=\epsilon (-i)^{n}\sum_{\alpha =-\infty }^{\infty }\exp \left[ -%
\frac{\pi }{N}\alpha ^{2}+\frac{2\pi i}{N}j\alpha \right] H_{n}\left(
\epsilon \alpha \right) ,  \label{def3}
\end{equation}%
which is in fact a generalization of Eq.(\ref{gaussum}) (if one compares it
to Eq.(\ref{def1}).

\section{Orthogonality of the $\ f_{n}{}^{\prime }s.$}

According to Eq.(\ref{mehta}), the functions $\{f_{n}(j)\}$ are eigenstates
of the DFT with associated eigenvalue $i^{n}.$ Mehta has conjectured that $%
\{f_{n}(j)\}_{n=0}^{N-1}$ is an orthogonal set, and thus complete, over a
finite set of $N$ points. This reasonable conjecture, quite surprisingly
indeed, does not hold for arbitrary $N$ (it holds for large $N$). As, in the
following, evidence will be collected \emph{against} the original
conjecture, details shall be kept to a level higher than usual.

Let $(f_{n},f_{m})$ denote the inner product
\begin{eqnarray*}
(f_{n},f_{m}) &=&\sum_{j=0}^{N-1}\left. f_{n}\right. ^{\ast }(j)f_{m}(j) \\
&=&\epsilon ^{2}(-i)^{n+m}\sum_{j=0}^{N-1}\sum_{\alpha ,\beta =-\infty
}^{\infty }\exp \left[ -\frac{\pi }{N}(\alpha ^{2}+\beta ^{2})+\frac{2\pi i}{%
N}j(\alpha -\beta )\right] H_{n}\left( \epsilon \alpha \right) H_{m}\left(
\epsilon \beta \right) .
\end{eqnarray*}%
The sum over $\{j\}$ is a realization of the modulo $N$ Kroenecker delta,
\begin{equation*}
\delta _{\alpha ,\beta }^{[N]}=\left\{
\begin{array}{cc}
1 & \alpha =\beta \ (\mbox{mod}N) \\
0 & \alpha \neq \beta \ (\mbox{mod}N)%
\end{array}%
\right.
\end{equation*}%
thus
\begin{equation}
(f_{n},f_{m})=2\pi (-i)^{n+m}\sum_{\alpha ,\beta =-\infty }^{\infty }\delta
_{\alpha ,\beta }^{[N]}\exp \left[ -\frac{\pi }{N}(\alpha ^{2}+\beta ^{2})%
\right] H_{n}\left( \epsilon \alpha \right) H_{m}\left( \epsilon \beta
\right) .  \label{kron}
\end{equation}%
The well known identity,
\begin{equation*}
\exp \left[ -\frac{1}{2}x^{2}\right] H_{k}(x)=\frac{i^{k}}{\sqrt{2\pi }}%
\int_{-\infty }^{\infty }dy\exp \left[ -\frac{1}{2}y^{2}+ixy\right] H_{k}(y),
\end{equation*}%
together with the sum over $\{\beta \}$ leads to
\begin{equation*}
(f_{n},f_{m})=\sum_{\alpha ,\gamma =-\infty }^{\infty }\int_{-\infty
}^{\infty }dydz\exp \left[ -\frac{1}{2}(y^{2}+z^{2})+iy\epsilon \alpha
+iz\epsilon \left( \alpha +\gamma N\right) \right] H_{n}(y)H_{m}(z),
\end{equation*}%
where the infinite sum on $\{\gamma \}$ covers the equivalence class present
in $\delta _{\alpha ,\beta }^{[N]}.$ Now, the sum over $\left\{ \alpha
\right\} $ by its turn is a realization of a modulo $2\pi $ Dirac delta,
thus, with the integration over $\{z\}$ and convenient changes of variables,
\begin{equation*}
(f_{n},f_{m})=2\pi \sum_{\gamma ,\upsilon =-\infty }^{\infty }\int_{-\infty
}^{\infty }dy\exp \left[ -y^{2}-\frac{\pi ^{2}v^{2}}{\epsilon ^{2}}%
+i(y\epsilon N-v\pi N)\gamma \right] H_{n}\left( y-\frac{\pi v}{\epsilon }%
\right) H_{m}\left( y+\frac{\pi \upsilon }{\epsilon }\right) .
\end{equation*}%
where again an infinite sum is introduced due to the modulo $2\pi $ delta.

The above expression is rather elucidative. It is not hard to realize that
the infinite sums over $\{\gamma ,\upsilon \}$ are a direct consequence of
the equivalence classes brought in by the modulo $N$ Kroenecker delta
present in Eq.(\ref{kron}). For large $N$, the term corresponding to $\gamma
=\upsilon =0$ becomes increasingly important, and a simple check shows that
this term is exactly $\delta _{n,m}$. Thus, as expected, the limit $%
N\rightarrow \infty $ recovers the usual harmonic oscillator results. For
finite (and small) $N$, however, all terms in the above summation must be
taken into account.

Following then, the sum on $\gamma $ is seen to be a realization of the
modulo $2\pi $ Dirac delta, $\delta ^{\lbrack 2\pi ]}(y\epsilon N-v\pi N)$,
and after a change of variables one has

\begin{equation*}
(f_{n},f_{m})=2\pi \epsilon \sum_{\mu ,\upsilon =-\infty }^{\infty }\exp
\left[ -\frac{2\pi }{N}\left( \mu +\frac{Nv}{2}\right) ^{2}-\frac{N\pi v^{2}%
}{2}\right] H_{n}\left( \epsilon \mu \right) H_{m}\left( \frac{2\pi v}{%
\epsilon }+\epsilon \mu \right) .
\end{equation*}%
Again, summation over on $\left\{ \mu \right\} $ must be included to account
for the $2\pi $ periodicity of the Dirac delta. Splitting the sum on $\nu $
in two sums, over the odd and even integers and shifting the sum on $\mu $
by $\nu N$ results in
\begin{eqnarray*}
(f_{n},f_{m}) &=&2\pi \epsilon \sum_{\mu ,\upsilon =-\infty }^{\infty }\exp
\left[ -\frac{2\pi }{N}\mu ^{2}-2\pi Nv^{2}\right] H_{n}(\epsilon \mu
-\epsilon Nv)H_{m}(\epsilon \mu +\epsilon Nv) \\
&&+2\pi \epsilon \sum_{\mu ,\upsilon =-\infty }^{\infty }\exp \left[ -\frac{%
2\pi }{N}(\mu +\frac{N}{2})^{2}-2\pi N(v+1/2)^{2}\right] H_{n}(\epsilon \mu
-\epsilon Nv)H_{m}(\epsilon (\mu +vN+N)).
\end{eqnarray*}%
Denoting the second term above by $(f_{n},f_{m})_{\text{odd}}$, if $N=2h+k$,
where the binary variable $k$ controls the parity of $N,$ then
\begin{equation*}
(f_{n},f_{m})_{\text{odd}}=2\pi \epsilon \sum_{\mu ,\upsilon =-\infty
}^{\infty }\exp \left[ -\frac{2\pi }{N}(\mu +h+k/2)^{2}-2\pi N(v+1/2)^{2}%
\right] H_{n}(\epsilon \mu -\epsilon Nv)H_{m}(\epsilon (\mu +vN+N))
\end{equation*}%
and yet again shifting the sum on $\mu $ by $h+k/2$ and the one on $\nu $ by
$1/2,$%
\begin{equation*}
(f_{n},f_{m})_{\text{odd}}=2\pi \epsilon \sum_{\mu =-\infty }^{\infty
}(k)\sum_{\upsilon =-\infty }^{\infty }(1)\exp \left[ -\frac{2\pi }{N}\mu
^{2}-2\pi Nv^{2}\right] H_{n}(\epsilon (\mu -Nv))H_{m}(\epsilon (\mu +vN)),
\end{equation*}%
where now $\sum_{\mu =-\infty }^{\infty }(k)$ denotes a sum over the
integers (half-integers) if $k=0$ ($k=1$), so that back to the general
expression,

\begin{eqnarray*}
(f_{n},f_{m}) &=&2\pi \epsilon \sum_{\mu ,\upsilon =-\infty }^{\infty }\exp
\left[ -\frac{2\pi }{N}\mu ^{2}-2\pi Nv^{2}\right] H_{n}(\epsilon (\mu
-Nv))H_{m}(\epsilon (\mu +vN)) \\
+ &&2\pi \epsilon \sum_{\mu =-\infty }^{\infty }(k)\sum_{\upsilon =-\infty
}^{\infty }(1)\exp \left[ -\frac{2\pi }{N}\mu ^{2}-2\pi Nv^{2}\right]
H_{n}(\epsilon (\mu -Nv))H_{m}(\epsilon (\mu +vN)).
\end{eqnarray*}%
Now, recourse to the Hermite polynomial's generating function gives
\begin{eqnarray*}
(f_{n},f_{m}) &=&2\pi \epsilon \frac{\partial ^{n}}{\partial t^{n}}\frac{%
\partial ^{m}}{\partial s^{m}}\left\{ \sum_{\mu ,\upsilon =-\infty }^{\infty
}\exp \left[ -\frac{2\pi }{N}\mu ^{2}+2\mu \epsilon (t+s)-2\epsilon \nu
N(t-s)-t^{2}-s^{2}\right] \right.  \\
&&+\left. \sum_{\mu =-\infty }^{\infty }(k)\sum_{\upsilon =-\infty }^{\infty
}(1)\exp \left[ -\frac{2\pi }{N}\mu ^{2}+2\mu \epsilon (t+s)-2\epsilon \nu
N(t-s)-t^{2}-s^{2}\right] \right\} _{t=s=0}.
\end{eqnarray*}%
The sum on $\{\mu \}$ results in a $\vartheta _{3}$-function in the first
term, and a $\vartheta _{3}$ for $k=0$ or a $\vartheta _{2}$ for $k=1$ in
the second. The sum on $\{\nu \}$, by its turn, gives $\vartheta _{3}$%
-function in the first term, and a $\vartheta _{2}$ in the second, as
\begin{eqnarray*}
(f_{n},f_{m}) &=&2\pi \epsilon \frac{\partial ^{n}}{\partial t^{n}}\frac{%
\partial ^{m}}{\partial s^{m}}\left\{ \left[ \vartheta _{3}\left( \frac{%
i\epsilon (t+s)}{\pi },\frac{2i}{N}\right) \vartheta _{3}\left( \frac{%
i\epsilon N(t-s)}{\pi },2Ni\right) \right. +\right.  \\
&&\left. \left. \left. \vartheta _{3-k}\left( \frac{i\epsilon (t+s)}{\pi },%
\frac{2i}{N}\right) \vartheta _{2}\left( \frac{i\epsilon N(t-s)}{\pi }%
,2Ni\right) \right] \left. \exp \left[ -t^{2}-s^{2}\right] \right\vert
\right\} \right\vert _{t=s=0}.
\end{eqnarray*}%
Using the basic properties,
\begin{eqnarray*}
\vartheta _{3}\left( z,i\tau \right)  &=&\tau ^{-1/2}\exp \left[ -\frac{\pi
z^{2}}{\tau }\right] \vartheta _{3}\left( \frac{z}{i\tau },\frac{i}{\tau }%
\right)  \\
\theta _{2}\left( z,i\tau \right)  &=&\tau ^{-1/2}\exp \left[ -\frac{\pi
z^{2}}{\tau }\right] \vartheta _{4}\left( \frac{z}{i\tau },\frac{i}{\tau }%
\right) ,
\end{eqnarray*}%
one gets
\begin{eqnarray*}
(f_{n},f_{m}) &=&\frac{2\pi ^{3/2}}{N}\frac{\partial ^{n}}{\partial t^{n}}%
\frac{\partial ^{m}}{\partial s^{m}}\left\{ \left[ \vartheta _{3}\left(
\frac{i\epsilon (t+s)}{\pi },\frac{2i}{N}\right) \vartheta _{3}\left( \frac{%
\epsilon (t-s)}{2\pi },\frac{i}{2N}\right) \right. \right.  \\
&&+\left. \left. \left. \vartheta _{3-k}\left( \frac{i\epsilon (t+s)}{\pi },%
\frac{2i}{N}\right) \vartheta _{4}\left( \frac{\epsilon (t-s)}{2\pi },\frac{i%
}{2N}\right) \right] \left. \exp \left[ -2ts\right] \right\vert \right\}
\right\vert _{t=s=0}.
\end{eqnarray*}%
Finally, compact expressions can be achieved with
\begin{eqnarray*}
\theta _{3}\left( z,\tau \right)  &=&\frac{1}{2}\left[ \theta _{3}\left(
\frac{z}{2},\frac{\tau }{4}\right) +\theta _{4}\left( \frac{z}{2},\frac{\tau
}{4}\right) \right]  \\
\theta _{2}\left( z,\tau \right)  &=&\frac{1}{2}\left[ \theta _{3}\left(
\frac{z}{2},\frac{\tau }{4}\right) -\theta _{4}\left( \frac{z}{2},\frac{\tau
}{4}\right) \right] ,
\end{eqnarray*}%
thus for $k=0$

\begin{equation*}
(f_{n},f_{m})=\frac{\pi ^{3/2}}{N}\frac{\partial ^{n}}{\partial t^{n}}\frac{%
\partial ^{m}}{\partial s^{m}}\left. \vartheta _{3}\left( \frac{i\epsilon
(t+s)}{\pi },\frac{2i}{N}\right) \vartheta _{3}\left( \frac{\epsilon (t-s)}{%
\pi },\frac{2i}{N}\right) \exp \left[ -2ts\right] \right\vert _{t=s=0}
\end{equation*}
and for $k=1$

\begin{eqnarray*}
(f_{n},f_{m}) &=&\frac{\pi ^{3/2}}{N}\frac{\partial ^{n}}{\partial t^{n}}%
\frac{\partial ^{m}}{\partial s^{m}}\left\{ \vartheta _{3}\left( \frac{%
i\epsilon (t+s)}{\pi },\frac{2i}{N}\right) \vartheta _{3}\left( \frac{%
\epsilon (t-s)}{\pi },\frac{2i}{N}\right) \right.  \\
&&\left. -2\vartheta _{4}\left( \frac{i\epsilon (t+s)}{2\pi },\frac{i}{2N}%
\right) \vartheta _{4}\left( \frac{\epsilon (t-s)}{2\pi },\frac{i}{2N}%
\right) \exp \left[ -2ts\right] \right\} _{t=s=0}.
\end{eqnarray*}%
Again, the limit $N\rightarrow \infty $ easily recovers the usual results,
as the $i$ factor inside the $\vartheta $-functions guarantees that, in this
limit, only a term proportional to $\frac{\partial ^{n}}{\partial t^{n}}%
\frac{\partial ^{m}}{\partial s^{m}}\left. \exp \left[ -4ts\right]
\right\vert _{t=s=0}$ $\ $survives. Anyhow, with the above expressions any
term $(f_{n},f_{m})$ can be calculated as a sum of $\vartheta $-function
derivatives evaluated at zero. The particular situation $m=0$, for example,
for $N$ even, is quite instructive. In this case
\begin{eqnarray*}
(f_{n},f_{0}) &=&\frac{\pi ^{3/2}}{N}\left. \frac{\partial ^{n}}{\partial
t^{n}}\vartheta _{3}\left( \frac{i\epsilon t}{\pi },\frac{2i}{N}\right)
\vartheta _{3}\left( \frac{\epsilon t}{\pi },\frac{2i}{N}\right) \right\vert
_{t=0} \\
(f_{n},f_{0}) &=&\frac{\pi ^{3/2}}{N}\sum_{j=0}^{n}\binom{n}{j}\left. i^{j}%
\frac{\partial ^{j}}{\partial t^{j}}\vartheta _{3}\left( \frac{\epsilon t}{%
\pi },\frac{2i}{N}\right) \right\vert _{t=0}\left( \left. \frac{\partial
^{n-j}}{\partial t^{n-j}}\vartheta _{3}\left( \frac{\epsilon t}{\pi },\frac{%
2i}{N}\right) \right\vert _{t=0}\right) ,
\end{eqnarray*}%
and its is immediate to see that all $n=$odd terms are zero. For $n=2$ (and
for all even numbers not multipliers of $4),$ the symmetry of the binomial
term and the multiplicity of the powers of $i$ lead to a pairwise
cancelation of all non-zero terms. For $n=4$ (and its multipliers), the
situation its different. The simplest case is $n=4$,
\begin{eqnarray*}
(f_{4},f_{0}) &=&\frac{\pi ^{3/2}}{N}\sum_{j=0}^{4}\binom{4}{j}\left. i^{j}%
\frac{\partial ^{j}}{\partial t^{j}}\vartheta _{3}\left( \frac{\epsilon t}{%
\pi },\frac{2i}{N}\right) \right\vert _{t=0}\left[ \left. \frac{\partial
^{4-j}}{\partial t^{4-j}}\vartheta _{3}\left( \frac{\epsilon t}{\pi },\frac{%
2i}{N}\right) \right\vert _{t=0}\right]  \\
(f_{4},f_{0}) &=&\frac{\pi ^{3/2}}{N}\left\{ 2\vartheta _{3}\left( 0,\frac{2i%
}{N}\right) \vartheta _{3}^{\prime \prime \prime \prime }\left( 0,\frac{2i}{N%
}\right) -6\left[ \vartheta _{3}^{\prime \prime }\left( 0,\frac{2i}{N}%
\right) \right] ^{2}\right\} .
\end{eqnarray*}%
This term (with proper normalization) goes to zero quite fast with
increasing $N$. In fact, for $N=10$ it is already of order of $10^{-6}$. On
the other hand, it is immaterial to discuss the case $N=4$ (or smaller), as
in this situation the distinct eigenvalues of the Fourier operator are
enough to guarantee orthogonality of the whole set. Considering all this, it
comes down to, literally, half a dozen different values of the
dimensionality $N$ (the range $[5,10]$) for which a significant deviation
from the \textquotedblleft expected\textquotedblright\ results (that is,
orthogonality) can be observed.

\section{DFT and width inversion}

Starting from the own definition of the $\vartheta _{3}$-function, Eq.(\ref%
{theta}), with $\xi \in \mathbf{R}$, a fractional shift of the $\vartheta
_{3}$ function can be calculated,
\begin{equation*}
\vartheta _{3}\left( z+\frac{k}{N},\frac{i\xi ^{2}}{N}\right) =\sum_{\alpha
=-\infty }^{\infty }\exp \left[ -\frac{\pi }{N}\xi ^{2}\alpha ^{2}\right]
\exp \left[ 2\pi i\alpha \left( z+\frac{k}{N}\right) \right] ,
\end{equation*}
where $k$ is an integer. The sum over $\{\alpha \}$ can be broken into
modulo $N$ equivalence classes as
\begin{equation*}
\vartheta _{3}\left( z+\frac{k}{N},\frac{i\xi ^{2}}{N}\right)
=\sum_{j=0}^{N}\sum_{\beta =-\infty }^{\infty }\exp \left[ -\frac{\pi }{N}%
\xi ^{2}(j+\beta N)^{2}\right] \exp \left[ 2\pi i(j+\beta N)\left( z+\frac{k%
}{N}\right) \right] .
\end{equation*}
Conveniently regrouping the terms one gets
\begin{eqnarray*}
\vartheta _{3}\left( z+\frac{k}{N},\frac{i\xi ^{2}}{N}\right)
&=&\sum_{j=0}^{N}\left( \sum_{\beta =-\infty }^{\infty }\exp \left[ -\frac{%
\pi }{N}\xi ^{2}\beta ^{2}\right] \exp \left[ 2\pi i\beta \left( i\xi
^{2}j+Nz\right) \right] \right) \\
&&\times \exp \left[ -\frac{\pi }{N}\xi ^{2}j^{2}+2\pi ijz+\frac{2\pi i}{N}jk%
\right] ,
\end{eqnarray*}
where the term inside the brackets can be identified as $\vartheta _{3}$%
-function,
\begin{equation*}
\vartheta _{3}\left( z+\frac{k}{N},\frac{i\xi ^{2}}{N}\right)
=\sum_{j=0}^{N}\vartheta _{3}\left( i\xi ^{2}j+Nz,iN\xi ^{2}\right) \exp %
\left[ -\frac{\pi }{N}\xi ^{2}j^{2}+2\pi ijz+\frac{2\pi i}{N}jk\right] .
\end{equation*}
Use of property (\ref{prop2}) leads to
\begin{equation}
\vartheta _{3}\left( z+\frac{k}{N},\frac{i\xi ^{2}}{N}\right) =\frac{1}{%
\sqrt{N\xi ^{2}}}\sum_{j=0}^{N}\vartheta _{3}\left( \frac{iz}{\xi ^{2}}-%
\frac{j}{N},\frac{i}{N\xi ^{2}}\right) \exp \left[ -\frac{\pi N}{\xi ^{2}}%
z^{2}+\frac{2\pi i}{N}jk\right]  \label{resultado1}
\end{equation}
and taking advantage of the Fourier coefficients $\exp \left[ \frac{2\pi i}{N%
}jk\right] $ it is easy to obtain the inverse relation
\begin{equation}
\vartheta _{3}\left( \frac{iz}{\xi ^{2}}-\frac{k}{N},\frac{i}{N\xi ^{2}}%
\right) =\sqrt{\frac{N}{\xi ^{2}}}\sum_{j=0}^{N}\vartheta _{3}\left( z+\frac{%
j}{N},\frac{i\xi ^{2}}{N}\right) \exp \left[ \frac{\pi N}{\xi ^{2}}z^{2}-%
\frac{2\pi i}{N}jk\right] .  \label{resultado2}
\end{equation}
Particular cases of these equations are most interesting, and a lot of
peculiar relations can be obtained with the different possible choices of $%
z,k$ and $\xi $. Two straightforward examples are: First, putting $z=0$ in (%
\ref{resultado1}),
\begin{equation}
\vartheta _{3}\left( \frac{k}{N},\frac{i\xi ^{2}}{N}\right) =\frac{1}{\sqrt{N%
}}\sum_{j=0}^{N}\vartheta _{3}\left( \frac{j}{N},\frac{i}{N\xi ^{2}}\right)
\exp \left[ \frac{2\pi i}{N}jk\right] ,  \label{fourier}
\end{equation}
and, according to Eq.(\ref{gaussum}), the $\vartheta _{3}$-function on the
left hand side has width $\xi ,$ while the one under the action of the DFT
has width $\xi ^{-1}$. This property is the obvious discrete counterpart of
the well known behavior of the Gaussian function under the usual Fourier
transform.

The case $k=0$, by its turn, after some manipulation gives
\begin{equation*}
\vartheta _{3}\left( Nz,iN\xi ^{2}\right) =\sqrt{\frac{N}{\xi ^{2}}}%
\sum_{j=0}^{N}\vartheta _{3}\left( z+\frac{j}{N},\frac{i\xi ^{2}}{N}\right) .
\end{equation*}

\subsection{\protect\bigskip Application}

With the above results it is possible to generalize the result of \cite%
{mehta} in a straightforward way. Introducing
\begin{equation*}
f_{n}(j,\xi )=\sqrt{\frac{N}{\xi }}\left. \frac{\partial ^{n}}{\partial t^{n}%
}\vartheta _{3}\left( \frac{j}{N}-\frac{\epsilon }{\pi }\xi t,\frac{i\xi ^{2}%
}{N}\right) \exp \left[ t^{2}\right] \right| _{t=0},
\end{equation*}
its DFT can be directly calculated,
\begin{eqnarray*}
\overline{f_{n}}(k,\xi ) &=&\frac{1}{\sqrt{N}}\sum_{j=0}^{N-1}\exp \left[
\frac{2\pi i}{N}jk\right] f_{n}(j,\xi ) \\
\overline{f_{n}}(k,\xi ) &=&\frac{1}{\sqrt{\xi }}\frac{\partial ^{n}}{%
\partial t^{n}}\sum_{j=0}^{N-1}\exp \left[ \frac{2\pi i}{N}jk\right] \left.
\vartheta _{3}\left( \frac{j}{N}-\frac{\epsilon }{\pi }\xi t,\frac{i\xi ^{2}%
}{N}\right) \exp \left[ t^{2}\right] \right| _{t=0},
\end{eqnarray*}
and use of Eq.(\ref{prop2}) together with change of variables from $t$ to $%
it $ leads to
\begin{equation*}
\overline{f_{n}}(k,\xi )=\sqrt{N\xi }i^{n}\frac{\partial ^{n}}{\partial t^{n}%
}\left. \vartheta _{3}\left( \frac{k}{N}-\frac{\epsilon t}{\pi \xi },\frac{i%
}{N\xi ^{2}}\right) \exp \left[ t^{2}\right] \right| _{t=0},
\end{equation*}
thus
\begin{equation}
f_{n}(k,\xi ^{-1})=i^{n}\sum_{j=0}^{N-1}\exp \left[ \frac{2\pi i}{N}jk\right]
f_{n}(j,\xi ),  \label{mg}
\end{equation}
with reproduces Eq.(\ref{mehta}) for $\xi =1.$ From this relation, most
identities obtained in \cite{mehta} may also be generalized.

\section{ Two variable's DFT}

Yet another generalization of the main result of \cite{mehta} regards a two
variable DFT, which, for the sake of briefness, here it will be merely
verified. Apart from the obvious product solution $f_{m}(j)f_{n}(l),$ if one
considers the quantity
\begin{equation*}
F_{m,n}(j,l)=\sum_{k=0}^{N-1}f_{m}(k)f_{n}(k-l)\exp \left[ \frac{2\pi i}{N}jk%
\right] ,
\end{equation*}
which obeys
\begin{equation*}
\left( F_{m,n}(j,l)\right) ^{\ast }=F_{m,n}(j,l)\exp \left[ \frac{2\pi i}{N}%
jl\right] ,
\end{equation*}
use of Eq.(\ref{mehta}), and some simple manipulations lead to the non
trivial result
\begin{equation*}
\left| F_{m,n}(j,l)\right| ^{2}=\frac{(-i)^{m+n}}{N}\sum_{a,b=0}^{N-1}\left|
F_{m,n}(a,b)\right| ^{2}\exp \left[ \frac{2\pi i}{N}(ma+nb)\right] .
\end{equation*}
As in the one variable case, this states obey
\begin{equation*}
\sum_{j,l=0}^{N-1}\left| F_{m,n}(j,l)\right| ^{2}\left| F_{m^{\prime
},n^{\prime }}(j,l)\right| ^{2}=\delta _{m,m^{\prime }}\delta _{n,n^{\prime
}}\qquad m+n\neq m^{\prime }+n^{\prime }(\mbox{mod}4),
\end{equation*}
which imply a multitude of relations involving derivatives of the $\vartheta
_{3}$-functions (or the Hermite polynomials). Motivated by the previous
section, it should be investigated wether this relations holds for $%
m+n=m^{\prime }+n^{\prime }(\mbox{mod}4).$

\section{Further relations involving the width}

So far it has been seen that to break up the infinite sum present in the
definition of the Jacobi $\vartheta _{3}$- function leads to interesting
properties of this very function. In order to further explore this
technique, from Eq.(\ref{gaussum}) it is straightforward to write
\begin{equation}
\vartheta _{3}\left( \frac{z}{L},\frac{i\xi ^{2}}{L}\right) =\frac{\sqrt{L}}{%
\xi }\sum_{\alpha =-\infty }^{\infty }\exp \left[ -L\pi \left( \frac{z}{\xi L%
}+\frac{\alpha }{\xi }\right) ^{2}\right] ,
\end{equation}
with $L$ a positive real number. Choosing $\xi $ integer, it is possible to
break the sum over $\{\alpha \}$ into modulo $\xi $ equivalence classes
\begin{eqnarray*}
\vartheta _{3}\left( \frac{z}{L},\frac{i\xi ^{2}}{L}\right) &=&\frac{\sqrt{L}%
}{\xi }\sum_{j=0}^{\xi -1}\sum_{\mu =-\infty }^{\infty }\exp \left[ -L\pi
\left( \frac{z}{\xi L}+\frac{j+\mu \xi }{\xi }\right) ^{2}\right] \\
\vartheta _{3}\left( \frac{z}{L},\frac{i\xi ^{2}}{L}\right) &=&\frac{\sqrt{L}%
}{\xi }\sum_{j=0}^{\xi -1}\sum_{\mu =-\infty }^{\infty }\exp \left[ -L\pi
\left( \frac{z+jL}{\xi L}+\mu \right) ^{2}\right]
\end{eqnarray*}
and the infinite sum can be identified as a $\vartheta _{3}$
\begin{equation}
\vartheta _{3}\left( \frac{z}{L},\frac{i\xi ^{2}}{L}\right) =\frac{1}{\xi }%
\sum_{j=0}^{\xi -1}\vartheta _{3}\left( \frac{z+jL}{\xi L},\frac{i}{L}%
\right) .  \label{transf}
\end{equation}
It is quite interesting to put $z=z\xi $ above and observe that
\begin{equation*}
\vartheta _{3}\left( \frac{z\xi }{L},\frac{i\xi ^{2}}{L}\right) =\frac{1}{%
\xi }\sum_{j=0}^{\xi -1}\vartheta _{3}\left( \frac{z}{L}+\frac{j}{\xi },%
\frac{i}{L}\right) ,
\end{equation*}
which, for the particular case $\xi =2$ gives the well known result:
\begin{eqnarray*}
\vartheta _{3}\left( \frac{2z}{L},\frac{4i}{L}\right) &=&\frac{1}{2}\left[
\vartheta _{3}\left( \frac{z}{L},\frac{i}{L}\right) +\vartheta _{3}\left(
\frac{z}{L}+\frac{1}{2},\frac{i}{L}\right) \right] \\
\vartheta _{3}\left( \frac{2z}{L},\frac{4i}{L}\right) &=&\frac{1}{2}\left[
\vartheta _{3}\left( \frac{z}{L},\frac{i}{L}\right) +\vartheta _{4}\left(
\frac{z}{L},\frac{i}{L}\right) \right]
\end{eqnarray*}
Similar reasoning would lead to the complementary relation
\begin{equation}
\vartheta _{3}\left( \frac{z}{L},\frac{i}{L}\right) =\frac{1}{\xi }%
\sum_{j=0}^{\xi -1}\vartheta _{3}\left( \frac{z}{\xi L}+\frac{j}{\xi },\frac{%
i}{L\xi ^{2}}\right) .  \label{transf2}
\end{equation}
And again, the particular case $\xi =2$ gives
\begin{equation*}
\vartheta _{3}\left( \frac{z}{L},\frac{i}{L}\right) =\frac{1}{2}\left[
\vartheta _{3}\left( \frac{z}{2L},\frac{i}{4L}\right) +\vartheta _{4}\left(
\frac{z}{2L},\frac{i}{4L}\right) \right] .
\end{equation*}
Equations (\ref{transf}) and (\ref{transf2}) can be combined to provide an
alternative width inversion relation
\begin{equation*}
\vartheta _{3}\left( \frac{z\xi }{L},\frac{i\xi ^{2}}{L}\right) =\frac{1}{%
\xi ^{2}}\sum_{j,j^{\prime }=0}^{\xi -1}\vartheta _{3}\left( \frac{z}{\xi L}+%
\frac{j^{\prime }}{\xi }+\frac{j}{\xi ^{2}},\frac{i}{L\xi ^{2}}\right) .
\end{equation*}

\section{Conclusions}

The results here presented seem to argue in favor of one basic point: The
Jacobi $\vartheta _{3}$-function, together with the DFT, plays, in finite
dimensional spaces, the same role played by the Gaussian function in
conjunction with the usual Fourier transform. Concerning quantum mechanics,
Schwinger has already noted that, if the families of eigenstates of two
different observables are connected \textit{via} DFT, then those observables
share a maximum degree of incompatibility \cite{schw}. In this connection,
the width inversion relation obeyed by the $f_{n}(j,\xi )$ functions
strongly suggests that one may able to construct, for finite dimensional
spaces, states which behavior resemble that of the continuous minimum
uncertainty states.

However, such a reasoning meets an important hindrance if one considers that
the orthogonality of the DFT 's eigenstates ultimately fails. It is a fact,
however, that with increasing $N$ it becomes, in a numerical sense, true,
and in this case the $N\rightarrow \infty $ limit is reached, as witty as it
may sound, somewhere near a dozen. This fact may illustrate a true finite
dimensional idiosyncrasy, or it might lead one to look for the possibility
of finding different sets of DFT's eigenstates, an issue which is matter of
current research.

\textbf{Acknowledgment:} This work is supported by FAPESP under contract
number 03/13488-0. The author is grateful to E.C da Silva and D. Galetti for a reading of the manuscript and to D. Nedel and A. Gadelha for pertinent suggestions.

\end{document}